# Surface Passivation in Empirical Tight Binding

Yu He, Yaohua Tan, Zhengping Jiang, Michael Povolotskyi, Gerhard Klimeck, Tillmann Kubis
Network for Computational Nanotechnology, Purdue University, Indiana 47907, USA
Email: heyuyhe@gmail.com

Empirical Tight Binding (TB) methods are widely used in atomistic device simulations. Existing TB methods to passivate dangling bonds fall into two categories: 1) Method that explicitly includes passivation atoms is limited to passivation with atoms and small molecules only. 2) Method that implicitly incorporates passivation does not distinguish passivation atom types. This work introduces an implicit passivation method that is applicable to any passivation scenario with appropriate parameters. This method is applied to a Si quantum well and a Si ultra-thin body transistor oxidized with $SiO_2$ in several oxidation configurations. Comparison with ab-initio results and experiments verifies the presented method. Oxidation configurations that severely hamper the transistor performance are identified. It is also shown that the commonly used implicit H atom passivation overestimates the transistor performance.

Critical dimensions of modern semiconductor devices are settled in the domain of few thousands of atoms. Resolving the geometries and material compositions of these small devices in high detail is essential to accurately predict the electronic device performance. In particular, the surface treatment gets increasingly important since the surface-to volume ratio increases with shrinking device dimensions. Theoretical device predictions require atomic resolutions of all device features and many band treatment of electrons as offered e.g. by the empirical tight binding (TB) method [1][2]. The empirical TB method has been successfully applied to electronic band structure [3]-[5] and non-equilibrium transport calculations in modern nanodevices [6]-[8]. Surface atoms in TB contribute dangling bonds which often result in surface states within the material's band gap. This issue can be resolved when those dangling bonds are coupled to passivation atoms such as e.g. hydrogen atoms [9]. The two common numerical passivation methods are to either explicitly include passivation atoms and their coupling to the surface atoms within the electronic Hamiltonian matrix [9], or to alter the orbital energies of the dangling bonds with a passivation potential [10]. The explicit inclusion of passivation atoms is a very general approach and applicable to any semiconductor surface. However, the rank of the Hamiltonian matrices can increase significantly with the explicit inclusion of passivation atoms and their orbital degrees of freedom. This increases the numerical load particularly for nanodevices with a high surface-to volume ratio. For the case of zincblende and diamond crystal structures, Lee et al. have shown in Ref. [10] how to implicitly passivate $sp^3$-hybridized dangling bonds with passivation potentials only. In this way, the passivation does not increase the rank of the Hamiltonian matrices and the numerical load stays the same. However, due to the assumed $sp^3$-hybridization the model of Ref. [10] considers only passivation of $s$ and $p$ orbitals. It is also restricted to $sp^3$-hybridized bonding symmetries and does not distinguish between different passivation atoms (such as hydrogen and oxygen passivation). These aspects become increasingly relevant for state of the art nanodevices.

This work introduces a method to passivate dangling bonds in TB for arbitrary crystal structures and hybridization symmetries. This method distinguishes passivation atoms, since it uses ab-initio results for different passivation atoms as fitting targets. Similar to the method of Lee et al., this method does neither increase the rank of the electronic Hamiltonian nor the numerical complexity of solving band structure or electronic transport. In the following sections, the method is introduced and it is shown that it agrees with the one of Lee et al. for specific passivation parameters. The method is then applied to the passivation of Si (100) dangling bonds with $SiO_2$ in three different oxidation configurations. These different configurations are assessed with respect to their impact on electronic properties and IV characteristics of a concrete ultrathin body (UTB) field effect transistor.

The electronic Hamiltonian in the present method follows the standard TB approach for all non-surface atoms [11]. The Hamiltonian of each surface atom $H_{SS}$ is setup as

$$H_{SS} = H_0 + \lambda_P I + \sum_{P=1}^{N_{db}} \Sigma_{SS,P} \qquad (1)$$

Here, $H_0$ is the original Hamiltonian without passivation, $N_{db}$ is the number of dangling bonds, $\lambda_P$ is a surface potential, and $\Sigma_{SS,P}$ is a self-energy due to the coupling to the passivation atom $P$, which is given as

$$\Sigma_{SS,P} = H_{SP}(\varepsilon - H_P)^{-1} H_{PS} \qquad (2)$$

Notice that Eq. (2) is inspired by the contact self-energies of the non-equilibrium Green's function (NEGF) method [12]. The inversion in Eq. (2) represents the surface Green's function of the passivation atom $P$. In NEGF, $\varepsilon$ represents the electronic energy. In this work, however, $\varepsilon$ is a constant fitting parameter. $H_P$ is the Hamiltonian of the passivation atom and $H_{SP}$ is the coupling Hamiltonian between the surface atom and the passivation atom. If dangling bonds of two different surface atoms $S$ and $S'$ couple to the same passivation atom $P$, the passivation self-energy has interatomic contributions

$$\Sigma_{SS',P} = H_{SP}(\varepsilon - H_P)^{-1} H_{PS'} \qquad (3)$$

Interatomic passivation self-energies beyond surface atoms that couple to the same passivation atom are ignored. All Hamiltonians $H_0$, $H_P$, $H_{SP}$ etc. are setup following the notation of Ref. [11]. All required TB parameters, i.e. the onsite orbital energies of the passivation atom and the interatomic interactions are determined by fitting the TB band structures to the HSE06 exchange correlation functional [13] results of VASP [14]. In the VASP calculations, PAW pseudopotentials [15] for the electron-ion interaction are considered. Three top most valence bands and three lowest conduction bands are considered as the fitting targets. The energy window for fitting is set as 1.2eV around the middle of the band gap. The TB parameters of Si atoms in this work are taken from Ref. [9]. The present model is implemented and all TB results of this work are solved with the nanodevice simulator NEMO5 [16]. To compare TB results with optical band gaps of experiments, the exciton binding energy is estimated following Ref. [17] and subtracted from the calculated band gap. This estimation of the optical band gap was successfully applied to 2D $MoS_2$ in Ref. [17]. Ballistic transport in this work is solved with the quantum transmitting boundary method (QTBM) [18] using NEMO5.

The presented passivation method is validated against the known passivation method of Ref. [10]. It reproduces the Hamiltonian of Ref. [10] with the parameters $\varepsilon$=1eV, $V_{ss\sigma}$=-2.739eV, $V_{sp\sigma}$=4.743eV, $E_s=V_{sd\sigma}=V_{ss*\sigma}=\lambda=0$. Given that only s-orbital parameters are needed to reproduce Ref. [10], one can interpret this known passivation method as a hydrogen passivation.

In the following it is exemplified on a 2.2nm thick Si (100) quantum well structure embedded in $SiO_2$ that a careful treatment of the passivation atom type is needed to realistically predict device performances. It is shown in Ref. [19] that $Si/SiO_2$ has several interface configurations. The three configurations that differ most in their bandstructures are depicted in Fig.1 (a). In all configurations, the Si dangling bonds are partially saturated with oxygen atoms (O1) of $β$-cristobalite $SiO_2$. Remaining dangling bonds are either passivated with a double-bonded oxygen (O2) atom (DBM), or with two hydrogen atoms (HGM). In the bridging oxygen model (BOM), the dangling Si atoms that are not oxidized with $SiO_2$ are replaced with oxygen (O3) atoms. The coupling of dangling bonds with O3 is again modeled with the self-energy of Eqs. (2) and (3). Ab-initio HSE06 calculations show that the quantum well bandstructures differ significantly for these three different oxidation configurations (Fig.1 (b-d)). The ab-initio bandstructures in Fig.1 are very well reproduced with $sp3d5s*$ TB calculations of NEMO5 with the parameters of Table I. It is worth to mention that the number of oxygen and hydrogen parameters in Table I is common for 10 band TB models [9]. The important fitting targets and their fit quality are listed in Table II. The energy $\varepsilon$ is for DBM $\varepsilon$=0.008eV, for BOM $\varepsilon$=0.02335eV, and for HGM $\varepsilon$=-0.02324eV.

The oxidation configurations DBM and BOM do not suppress surface states completely, but host significant electronic density at the O2 and O3 atoms and Si atoms coupled to them. This agrees with findings of Ref. [20]. Such a surface density of states (DOS) introduces trapped states at the $Si/SiO_2$ interface which is expected to weaken the gate control of the transistor. Therefore, it is advisable to avoid these configurations in transistors. In contrast, the HGM configuration suppresses surface states very well, similar to the pure H atom passivation of Ref. [9].

Since each O atom contributes eight orbitals (with six occupied) and each H atom only two (with one occupied), the device DOS should be the larger, the fewer H are used for passivation. Oxidation processes that only add O atoms (DBM) should have a higher DOS than cases that replace some Si atoms with O atoms (BOM). This is confirmed in Fig.2 which shows the DOS in the 2.2nm Si quantum well passivated in different ways. Figure 2 (a) shows the results of pure H passivation following Ref. [10] and the HGM configuration of this work. The inset in Fig. 2(a) emphasizes the DOS of HGM is larger than the one of the pure H passivation for energies above the conduction band edge. Figure 2 (b) shows the DOS of HGM is exceeded by BOM and even more by DBM results. It can also be seen in Fig.2 that the band gap of pure H passivation agrees with the HGM model, but DBM and BOM results deviate from that. This is elaborated in Fig.3 which shows the thickness dependence of the Si quantum well optical band gaps solved in the TB oxidation models of this work. Experimental data of Ref. [21] are also shown for comparison. The calculated exciton binding energies for a 2.2nm Si quantum well in HGM and BOM configuration are 51meV and 75meV, respectively. This is of the same order as the exciton binding energy of homogeneous Si (20meV) [21]. For the DBM configuration, however, the exciton energy is 0.2eV due to its large effective masses (see Table II). Tight binding calculations with the HGM model reproduce the measured optical band gaps of Ref. [21] very well, while the TB results in the DBM and BOM configurations are much lower. The small variation of DBM band gaps with the quantum well thickness in Fig. 3 agrees with ab-initio results of Ref. [19].

Figure 4 shows the ballistic $I_d$-$V_g$ characteristics of a Si UTB transistor with H atom passivation and oxidation with $SiO_2$ in the HGM configuration. The Si UTB of Fig.4 follows the high performance logic technology requirements of ITRS 2020 [22]. The doping profile resembles an *n-i-n* UTB structure with $1.5 \times 10^{20}$cm$^{-3}$ in the *n* type doped regions. The thickness of the UTB is 3.3nm. The center 8.5nm long intrinsic channel is covered with a 2.8nm thick oxide layers on each UTB facet. Remaining UTB surfaces are passivated with H atoms following Ref. [10]. The threshold voltage $V_{th}$ is defined at $I_{off}=10^{-10}$A/nm. The ON-state current is defined at $V_g$-$V_{th}$=0.75V, and the source-drain bias $V_{ds}$=0.75V. All these values agree with the ITRS 2020 requirements [22]. The subthreshold slope resulting from the HGM model is 147mV/dec, which exceeds the 121mV/dec predicted in the pure H atom passivation model of Ref. [10] (see Fig.4). The ON/OFF ratio of the HGM model is $5.4 \times 10^3$ which is below the $7.7 \times 10^3$ of pure H atom passivation case. In conclusion, Si UTB transport calculations that model all dangling bonds passivated with only H atoms (following Ref. [10]) overestimate the transistor performance compared to calculations that consider gate areas covered with the best performing $SiO_2$.

In summary, this work introduces tight binding models for dangling bond passivation with $SiO_2$ in all relevant configurations. Ab-initio calculations served as input for fitting the passivation parameters. These models agree with an established H passivation model for a given parameter set. Tight binding band structure results of this work suggest that two of the $SiO_2$ configurations should be avoided in transistors due to adverse impact on the performance. It is also shown that passivation of all dangling bonds with only H atoms tends to overestimate the transistor performance.

Support by the SRC task 2141, SRC task 2273, by nanohub.org, and by the U.S. NSF (Nos. EEC-0228390, OCI-0832623 and OCI-0749140) are acknowledged.

TABLE I Passivation parameters in units of eV.

|  | O1 | O2 | O3 | H |
|---|---|---|---|---|
| $E_s$ | -0.0232 | 0.000548 | 0.00586 | -0.1232 |
| $E_p$ | 6.0978 | 19.774 | 0.23768 | NA |
| $E_{s*}$ | 1.7973 | 0.75406 | 0.00756 | |
| $E_d$ | 3.0662 | 5.0317 | 8.4569 | |
| $V_{ss\sigma}$ | -12.542 | -0.20058 | -0.1595 | -7.8087 |
| $V_{sp\sigma}$ | 0.00887 | -0.208 | 0.23644 | 9.3511 |
| $V_{sd\sigma}$ | 9.6611 | -0.19985 | -0.1992 | 0.076798 |
| $V_{ss*\sigma}$ | 1.8944 | -0.31368 | -0.00116 | 0.18137 |
| $V_{ps*\sigma}$ | 3.2026 | 0.17488 | -0.07399 | NA |
| $V_{pp\sigma}$ | -0.1426 | 0.01789 | 0.62144 | |
| $V_{pp\pi}$ | 0.2521 | -8.3849 | -0.36609 | |
| $V_{pd\sigma}$ | 0.1188 | 0.71363 | -1.7428 | |
| $V_{pd\pi}$ | 0.9182 | 0.75347 | 0.19001 | |
| $V_{dd\sigma}$ | 0.53801 | 9.26655 | -6.1452 | |
| $V_{dd\pi}$ | 1.3329 | 7.07982 | -0.37154 | |
| $V_{dd\delta}$ | 1.83785 | 8.510863 | 2.54124 | |
| $V_{s*s*\sigma}$ | 1.64669 | -2.96048 | -0.99583 | |
| $V_{s*d\sigma}$ | 0.01886 | -0.00109 | 0.12976 | |
| $\lambda$ | 3.05833 | 5.430484 | 0.60796 | -0.26006 |

TABLE II Tight binding (TB) and ab-initio (DFT) band gaps $E_g$, valence $E_v$ and conduction $E_c$ band edges in eV, and effective masses for electrons $m_e$ and holes $m_h$ at the $\Gamma$ point along $X$ and $M$ direction for the oxidation configurations of Fig.1.

|  | DBM | | BOM | | HGM | |
|---|---|---|---|---|---|---|
|  | DFT | TB | DFT | TB | DFT | TB |
| $E_v$ | -0.052 | -0.060 | -0.144 | -0.138 | -0.144 | -0.115 |
| $E_c$ | 1.12 | 1.115 | 1.111 | 1.121 | 1.237 | 1.22 |
| $E_g$ | 1.172 | 1.175 | 1.255 | 1.259 | 1.381 | 1.335 |
| $m_{e\ X}$ | 1.701 | 1.644 | 0.410 | 0.388 | 0.193 | 0.183 |
| $m_{e\ M}$ | 1.941 | 1.927 | 0.665 | 0.672 | 0.2 | 0.226 |
| $m_{h\ X}$ | -0.677 | -0.665 | -0.358 | -0.312 | -0.351 | -0.341 |
| $m_{h\ M}$ | -2.545 | -2.497 | -0.421 | -0.408 | -0.55 | -0.518 |

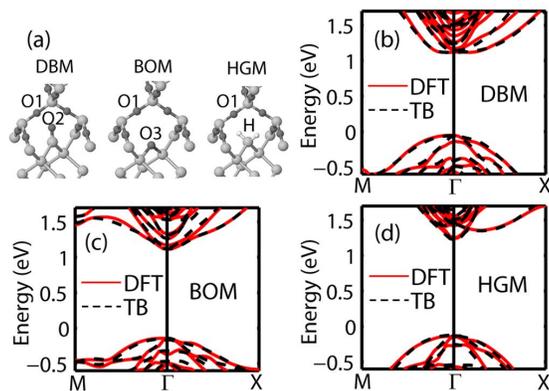

Figure 1 (a) Atomic structure of the three Si/SiO$_2$ configurations described in the main text. Spheres represent Si (dark gray), oxygen atoms with different surrounding O1, O2 and O3 (black), and H (light gray). The dispersion relations of ab-initio (solid) and TB (dashed) calculations of a 2.2nm thick Si quantum well oxidized in (b) DBM, (c) BOM, and (d) HGM configuration agree well.

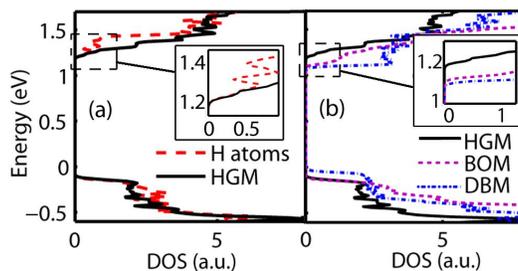

Figure 2 Density of states comparison of a 2.2nm Si quantum well for (a) HGM vs implicit H passivation, and (b) the three Si/SiO$_2$ configurations of Fig.1.

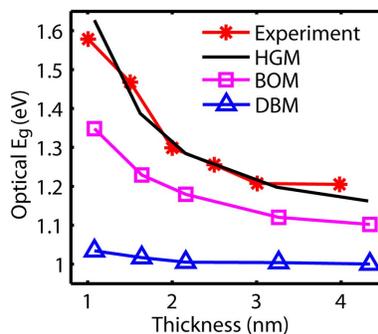

Figure 3 Calculated optical band gap of Si (100) quantum wells with varying thicknesses oxidized in the three configurations of Fig.1. Experimental data of Ref. [21] (asterisks) are given for comparison.

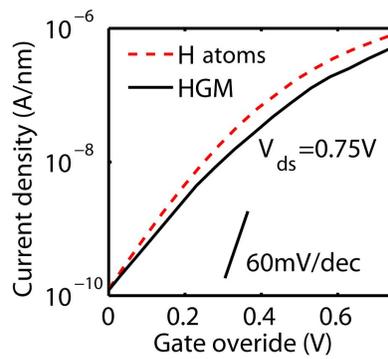

Figure 4 $I_d$-$V_g$ characteristic of the 3.3nm Si UTB transistor described in the main text when dangling bonds are passivated implicitly with H atoms following Ref. [10] (dashed) and with the present method for the HGM (solid) oxidization configuration.